\documentclass[aps,prb,amsmath,amssymb,
 reprint,%
]{revtex4-1}

\usepackage{graphicx}
\usepackage{dcolumn}
\usepackage{bm}

\begin{document}

\preprint{}

\title{Electronic structures and optical properties of realistic
transition metal dichalcogenide heterostructures from first principles}

\author{Hannu-Pekka Komsa$^1$}
\author{Arkady V. Krasheninnikov$^{1,2}$}
\affiliation{
$^1$Department of Physics, University of Helsinki,
P.O. Box 43, 00014 Helsinki, Finland
}
\affiliation{
$^2$Department of Applied Physics, Aalto University,
P.O. Box 11100, 00076 Aalto, Finland
}

\date{\today}

\begin{abstract}
We calculate from first principles the 
electronic structure and optical properties of a number of
transition metal dichalcogenide (TMD) bilayer heterostructures
consisting of MoS$_2$ layers sandwiched with
WS$_2$, MoSe$_2$, MoTe$_2$, BN, or graphene sheets.
Contrary to previous works, the systems are constructed in 
such a way that the unstrained lattice constants of the constituent 
incommensurate monolayers are retained. 
We find strong interaction between the $\Gamma$-point states in all
TMD/TMD heterostructures, which can lead to an indirect gap.
On the other hand, states near the K-point remain as in the monolayers.
When TMDs are paired with BN or graphene layers, the interaction 
around $\Gamma$-point is negligible, and the electronic
structure resembles that of two independent monolayers.
Calculations of optical properties of the MoS$_2$/WS$_2$ system 
show that even when the valence and conduction band edges are located 
in different layers, the mixing of optical transitions is minimal, and 
the optical characteristics of the monolayers are largely 
retained in these heterostructures. 
The intensity of interlayer transitions is found to be negligibly
small, a discouraging result for engineering the 
optical gap of TMDs by heterostructuring.

\end{abstract}


\maketitle

\section{Introduction}

Transition metal dichalcogenide (TMD) layered materials 
\cite{Wilson69_AP,Tenne92_Nat}
possess unique electronic,\cite{Radisavljevic11_NNano}
optical,\cite{Mak10_PRL, Xiao12_PRL, Mak13_NMat}
and mechanical properties.
\cite{KaplanAshiri06_PNAS, Bertolazzi11_ACSNano, Pu12_NL}
Moreover, when used together with graphene and BN sheets,
they show promise for construction of ultra-thin flexible
devices based solely on two-dimensional layers.
\cite{Novoselov12_PS,Wang12_NNano}
First steps in this direction have recently been demonstrated
with the fabrication of transistors, inverters, and memory cells.
\cite{Britnell12_Sci, Yu13_NMat, Bertolazzi13_ACSNano, Georgiou13_NNano}

In order to further expand the range of properties achievable 
by the TMD materials, the pristine systems may be modified through
doping \cite{Komsa12_PRL,Komsa13_PRB,Yadgarov12_ACIE,Cheng13_PRB} or
alloying,\cite{Komsa12_JPCL,Kang13_JAP,Dumcenco13_NComm}
and also by modifying the layer stacking.
Considering the dramatic change from indirect gap in bilayer MoS$_2$
to direct gap in monolayer,\cite{Mak10_PRL,Kuc11_PRB}
significant changes might also be expected when layers of
different materials are stacked.
For instance, recent computational studies suggested 
that the band gap may be engineered by constructing TMD/TMD 
heterostructures.\cite{Kosmider13_PRB,Terrones13_SRep,Kou13_JPCL}
However, several questions still remain open.

First, in the case of MoS$_2$/WS$_2$, it was shown that the valence 
band maximum (VBM) and conduction band minimum (CBM) are located 
in different layers (often referred to as ``type-II alignment'').
Through a simple picture of single-particle band structure,
such band alignment would lead to decreased band gaps. However, 
this is true only for the fundamental gap measured as a difference
between the electron affinity and ionization potential,
but does not necessarily hold for the optical transitions.
In essence, it is not clear what kind of transition spectrum
such systems would show.

Second, theoretical studies of other TMD/TMD heterostructures
(such as MoS$_2$/MoSe$_2$) suffer from the problem of incommensurate 
lattice constants, 
which has so far been circumvented by only considering strongly 
strained systems. The energy cost for straining both layers and 
the lack of barrier for relaxing to unstrained state suggest
that formation of strained heterostructures is unlikely.
Indeed, during epitaxial growth of TMD monolayers on
graphene, the lattice constants were found to be very close to 
those of isolated monolayers, although with some preferential 
orientation among the layers.\cite{Shi12_NL}
As strain is known to give rise to dramatic changes in the electronic
properties of 2D systems,
\cite{Yun12_PRB,Johari12_ACSNano,Shi13_PRB,Novoselov12_PS}
it is difficult to distinguish which features then originate from
the stacking and which are due to the strain artificially introduced
into the system due to computational limitations.

In this work, by
using first-principles calculations, we study bilayer 
heterostructures consisting of MoS$_2$ and WS$_2$, MoSe$_2$, 
MoTe$_2$, BN, or graphene. 
By constructing heterostructure models where both constituent layers 
retain their optimized lattice constant and including the electron-hole 
interactions through the solution of the Bether-Salpeter equation (BSE) for 
the MoS$_2$/WS$_2$ system, we go beyond the body of previous work and give 
answers to the open questions listed above.
We also discuss 
the optical properties of the systems with incommensurate lattices, and
demonstrate that the optical characteristics 
of the monolayers are largely retained in these heterostructures.

\section{Methods}

All calculations are carried out with plane waves and the
projector-augmented wave scheme as implemented in VASP.\cite{kres1,kres2}
The plane wave cutoff is set to 500 eV.
Exchange-correlation contributions are treated with
Perdew-Burke-Ernzerhof\cite{Perdew96} (PBE) functional including 
empirical dispersion corrections (PBE-D) proposed by Grimme.\cite{Grimme06}
The ions are relaxed until forces are converged to less than 2 meV/{\AA}.
In selected cases, we also crosscheck our PBE-D results against
ab-initio dispersion corrected functionals, PW86R-VV10 and AM05-VV10sol,
which have been shown to give layer distances in excellent agreement
with experiment.\cite{Vydrov10_JCP,Bjorkman12_PRB,Bjorkman12_PRL}

\begin{figure}[!ht]
\begin{center}
  \includegraphics[width=7.5cm]{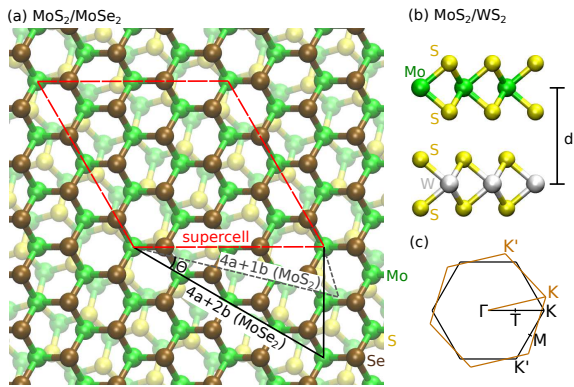}
\end{center}
\caption{\label{fig:geo}
(Color online)
(a) Top view of MoS$_2$/MoSe$_2$ bilayer heterostructure as
modeled in our supercell approach. The construction of the
supercell basis vectors ($n_ia_i+m_ib_i$) is also illustrated.
(b) Side view of the MoS$_2$/WS$_2$ heterostructure showing
the adopted stacking similar to 2H-polytype of MoS$_2$.
The definition for layer distance $d$ is indicated.
(c) Schematic demonstration of the overlap of the primitive
cell Brillouin zones from the MoS$_2$/MoSe$_2$ system.
}
\end{figure}

Since the constituent monolayers have generally differing lattice 
constants, special care is needed in the construction of
the atomic models in such a way that the strain is minimized.
Our approach is illustrated in Fig.~\ref{fig:geo}.
Let us denote the primitive cell basis vectors of a 2D material $i$
as $\{a_i,b_i\}$. The supercell basis vector may be 
constructed as $n_i a_i + m_i b_i$, where $n_i$ and $m_i$ are integers.
The second basis vector is always oriented at an 120 $^\circ$ angle.
We then search for a set of integers such that the magnitude of
the supercell basis vectors in materials $i$ and $j$ approximately
match: $|n_i a_i + m_i b_i| \approx |n_j a_j + m_j b_j|$.
In practice, we choose the smallest supercell for which the
strain is less than 1 \%. The resulting structures contain
75--102 atoms. The two integers defining our models are 
listed in Table \ref{tab:models}.
For all the heterostructures in this work, one layer is
always MoS$_2$. Therefore, we have fixed the lattice constant
of MoS$_2$ to the optimized value of 3.18 {\AA}, and
squeezed or stretched the other layer slightly.
These resulting lattice constants (compared to the optimized ones)
are also given in Table \ref{tab:models} together with the relative 
orientation  of the layers as determined by the construction scheme.
The perpendicular lattice vector $c$ is 25.44 {\AA}.
We employ 4$\times$4$\times$1 k-point sampling throughout.

\begin{table}[!t]
\begin{center}
\caption{\label{tab:models}
Description of supercell models for the bilayer heterostructures.
The integers describing the supercell basis vectors for both
layers are given together with the resulting lattice constant
of the second slightly strained layer (optimized lattice constant
in parentheses) and the angle between the lattices.
}
\begin{tabular}{lcccc}
system           & basis 1 & basis 2 & $a_2$ & angle \\
\hline
MoS$_2$/WS$_2$   & $a$     & $a$     & 3.18 (3.18) & 60.0 \\
MoS$_2$/MoSe$_2$ & $4a+1b$ & $4a+2b$ & 3.31 (3.32) & 16.1 \\
MoS$_2$/MoTe$_2$ & $4a$    & $4a+1b$ & 3.53 (3.55) & 13.9 \\
MoS$_2$/BN       & $4a+1b$ & $5a+1b$ & 2.50 (2.51) & 3.0 \\
MoS$_2$/G        & $4a$    & $6a+3b$ & 2.45 (2.47) & 30.0 \\
\end{tabular}
\end{center}
\end{table}

The calculated, as well as experimental, lattice constants of 
MoS$_2$ and WS$_2$ are very close.
Thus, the heterostructure can be constructed simply from the
primitive cells of MoS$_2$ and WS$_2$ with negligible strain.
The stacking adopted here corresponds to that found in 2H-MoS$_2$
i.e., chalcogen sublattice of one layer overlaps with
transition metal sublattice of the other layer.
Due to the large spin-orbit splitting of VBM states near
K-point, spin-orbit coupling is included in these calculations.
In this case, 12$\times$12$\times$1 k-point sampling is used.

The band structures are calculated for all systems
at the PBE level. The electronic states from the supercell 
calculations are projected to the primitive cells of 
each constituent layer, following Ref.~\onlinecite{Popescu10_PRL}.
Note, that when the layer orientations do not align, also the Brillouin
zone high-symmetry points for the two constituent monolayers reside
at different points of the reciprocal space, as illustrated
in Fig.~\ref{fig:geo}(c). 
When drawing the band structures, the Brillouin zone segments 
(e.g. $\Gamma$-K) from the two lattices are overlaid.
To be more precise, each supercell state is first projected to one
of the monolayers and then to the specific Brillouin zone segment 
of the respective layer.
Although the quasiparticle gap from $GW$ 
is strongly influenced by the size of vacuum, all the main features 
of the band structure are correctly described by PBE.\cite{Komsa12_PRB2}
Therefore, when the effects of interlayer interactions on the electronic 
structure are considered, the PBE level of theory is deemed sufficient.

Electron-hole interactions need to be accounted for when
optical absorption spectra are considered.
In the calculations of optical spectra, we rely on the
single-shot $G_0W_0$ procedure together with solution of
the Bethe-Salpeter equation in the Tamm-Dancoff 
approximation.\cite{Shishkin06_PRB,Rohlfing98_PRL}
Due to computational cost involved, this is only done for 
the MoS$_2$/WS$_2$ system.
Hybrid functionals are known to improve the starting 
electronic structure,\cite{Rinke05_NJP,Bechstedt09_PSSb}
and thus the $G_0W_0$ is solved on top of HSE band structure.
Atomic geometry is obtained using PBE-D with lattice constant
$c=18.67$ {\AA}, which leads to Mo-W distance of $d=12.49$ {\AA}
over the vacuum region. Note that unlike the $GW$ gaps,
the optical transitions from BSE are fairly insensitive to 
the amount of vacuum in the calculation.\cite{Wirtz06_PRL,Komsa12_PRB2}
Spin-orbit interaction is fully accounted for.
The adopted 12$\times$12$\times$1 k-point mesh is mostly sufficient
for proper description of excitons.\cite{Shi13_PRB,MolinaSanchez13_PRB}
For monolayer, we have 192 states in the calculation and for 
bilayer 288 states (scaled with the supercell volume).
Plane wave cutoffs are 280 eV and 200 eV for the wave functions
and for the response functions, respectively. 
These parameters were carefully optimized to reproduce
converged absorption spectrum.

\section{M\lowercase{o}S$_2$/WS$_2$ heterostructure}

\subsection{Electronic structure}

\begin{figure}[!ht]
\begin{center}
  \includegraphics[width=7.5cm]{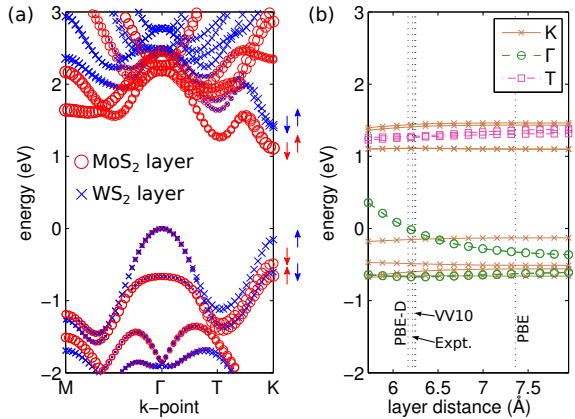}
\end{center}
\caption{\label{fig:MoSWSbands}
(Color online)
(a) Band structure of MoS$_2$/WS$_2$ heterostructure.
Projection to MoS$_2$ layer is denoted by red (circles)
and to WS$_2$ by blue (crosses).
The spin orientations of the wave functions at the K/K'-point 
are also denoted.
(b) The energies for the band edge states as a function of 
the interlayer distance $d$ [cf.~Fig.~\ref{fig:geo}(b)].
The vertical dotted lines denote distances calculated with various
functionals (the two VV10-type functionals give nearly identical results)
and experimental value evaluated from the average of bulk MoS$_2$ and WS$_2$.
}
\end{figure}

We first consider the MoS$_2$/WS$_2$ heterostructure, for which
the geometry and stacking is shown in Fig.~\ref{fig:geo}(b). 
The band structure in Fig.~\ref{fig:MoSWSbands}(a) shows the 
same main features as reported in 
Refs.~\onlinecite{Kosmider13_PRB,Terrones13_SRep}. 
In addition, the localization of the states to the two constituent 
monolayers are highlighted.
Around $\Gamma$-point, the VBM states show appreciable weight
in both layers. On the other hand, the states around K-point are 
stricly localized to one of the monolayers:
VBM at the K-point is completely localized to WS$_2$ and CBM to MoS$_2$.
The alignment is similar to that expected from the ionization potentials,
\cite{Jiang12_JPCC,Kang13_APL} indicating that the states also retain 
their energy position with respect to vacuum level.
The mixing around the $\Gamma$-point is due to interaction of the
constituent monolayer states, which also leads to strong shift
(or split) of the energy levels. As a result, 
the valence band $\Gamma$-point states are pushed 0.15 eV higher 
than the K-point states, thus making the gap indirect.

\begin{figure}[!ht]
\begin{center}
  \includegraphics[width=7.5cm]{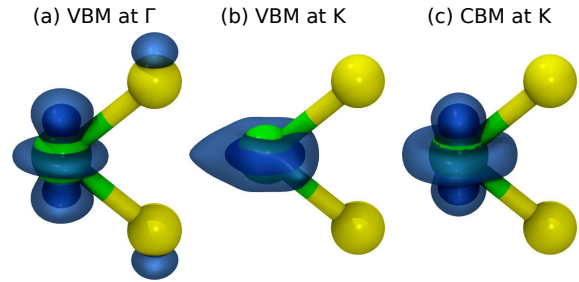}
\end{center}
\caption{\label{fig:wf}
(Color online)
Partial charge density isosurfaces (blue) 
at $0.37$ (transparent) and $1.2$ (solid) e/nm$^3$
for selected wave functions of monolayer MoS$_2$.
}
\end{figure}

The sensitivity of the VBM state at the $\Gamma$-point to the 
interlayer interactions may be immediately understood through
inspection of the constituent wave functions.
The wave functions at the valence and conduction band edges
are shown in Fig.~\ref{fig:wf} for MoS$_2$. They are very similar
for other TMDs. At the K-point, both for the VBM and the CBM,
wave functions are localized within the transition metal sublattice.
On the contrary, VBM at the $\Gamma$-point shows lobes extending 
out from the sulfur atoms. These states will interact strongly
(if also energetically close), when TMD layers are brought in contact.
This is further illustrated in Fig.~\ref{fig:MoSWSbands}(b),
where the band edge positions are plotted as a function of the
layer separation $d$. The $\Gamma$-point states are seen to move
strongly as the layers are brought in contact, while the states
around K- and T-valleys remain largely unaffected.
The gap becomes indirect at $d<6.47$ {\AA}, which is fulfilled
for all considered vdW-corrected functionals.
The layer distance calculated with the PBE-D, PW86R-VV10, and AM05-VV10sol 
functionals are $d=6.17$, $d=6.25$, and $d=6.24$ {\AA}, respectively.
PBE shows essentially no binding and gives the minimum at $d=7.36$ {\AA}.
Note that the experimental distances in both bulk MoS$_2$ and 
in bulk WS$_2$ are similar and would yield approximately $d=6.22$ {\AA},
in excellent agreement with the results calculated using the 
vdW-corrected functionals.

\subsection{Optical properties}

It was suggested previously,
that due to the type-II alignment of band edges in many of these
heterostructures, the optical band gap would also decrease
and that the excitons would have electron and hole localized
to different layers.\cite{Kosmider13_PRB,Terrones13_SRep,Kou13_JPCL}
In order to see if this is indeed the case, we calculated the 
optical absorption spectrum by solving the Bethe-Salpeter equation.
Since the simulation cell for MoS$_2$/WS$_2$ heterostructure is small,
computationally heavy $GW$+BSE calculations can be performed. 
In addition, the orientation of layers is likely to be ``correct'' 
in a sense that it corresponds to the minimum energy configuration.

\begin{figure}[!ht]
\begin{center}
  \includegraphics[width=7.5cm]{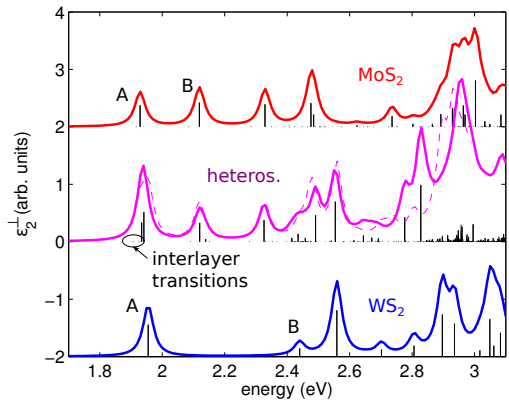}
\end{center}
\caption{\label{fig:abso}
(Color online) 
The optical absorption spectrum of MoS$_2$/WS$_2$ heterostructure
(middle) together with the spectra from 
monolayer MoS$_2$ (top) and WS$_2$ (bottom).
The vertical lines denote the actual calculated transition energies
and intensities. Absorption spectra are evaluated through application
of 0.02 eV Lorentzian broadening.
Overlaid with the explicitly calculated heterostructure spectrum,
the sum of the monolayer spectra is also plotted (dashed line).
Interlayer transitions are present, but have negligible intensities.
}
\end{figure}

As seen above, the VBM of this system is located at 
the $\Gamma$-point. However, the indirect transitions
do not show up in the absorption spectrum, for which the direct 
transitions at around K-point are known to 
dominate,\cite{Zhao13_ACSNano,Mak10_PRL,Komsa12_PRB2}
and we thus concentrate on the latter.
Nevertheless, at K-point the VBM is localized in the WS$_2$ layer 
and the CBM in the MoS$_2$ layer. The calculated absorption spectra 
for MoS$_2$ and WS$_2$ monolayers and for the MoS$_2$/WS$_2$ 
heterostructure are shown in Fig.~\ref{fig:abso}. 
We first note that the calculated energies for the lowest 
A/B transitions in the monolayer systems
(1.93/2.12 eV for MoS$_2$ and 1.96/2.44 eV for WS$_2$)
are in good agreement with the experimental ones
(1.9/2.1 eV for MoS$_2$ \cite{Mak12_NNano} and
2.0/2.4 eV for WS$_2$ \cite{Zhao13_ACSNano,Gutierrez13_NL}).
The optical response from MoS$_2$ and WS$_2$ monolayers are of
roughly equal intensity.
Turning now to the optical spectrum of the heterostructure, it
appears to hold all the same features as in the monolayers and
at the same energies. In fact, the total spectrum can be well 
approximated by simply summing up the monolayer spectra 
as shown by dashed line in Fig.~\ref{fig:abso}.
In order to analyze the character of each transition $S$,
we have inspected in more detail the electron-hole amplitude 
matrix $A_{vc{\bf k}}^S$,\cite{Rohlfing98_PRL}
where $v$, $c$, and ${\bf k}$ index valence band states,
conduction band states, and k-points, respectively.
This analysis shows that all optically active transitions
are comprised of direct intralayer transitions.
Note that with the adopted stacking, K-point of MoS$_2$ coincides
with K$'$ point of WS$_2$ and thus the spin orientations
of the MoS$_2$ and WS$_2$ VBM states are opposite, as shown
in Fig. \ref{fig:MoSWSbands}(a),
Interlayer transitions were also found, but their intensities 
are close to zero, and thus do not contribute to the absorption
spectrum in Fig.~\ref{fig:abso}.
These transitions reach only about 50 meV below the A peak, which 
is clearly less than that expected from the band alignment.
Due to spatial separation, the binding energy becomes smaller,
but relaxation of optically active intralayer exciton to 
optically inactive interlayer exciton is possible.

Thus, engineering of the optical band gap by stacking up suitably 
aligned monolayers, as previously suggested, does not appear
to work. On the other hand, it allows one to achieve stronger 
monolayer-like optical absorption from the layered TMD materials
by simply stacking them together. While our calculations do not give
the dynamics of various scattering and recombination mechanisms,
it is surely possible that the photoluminescence spectrum
could be modified by the interlayer excitons. 
Relaxation to interlayer excitons could be particularly useful in
separating and collecting optically excited electron-hole
pairs, and at the same time eliminating the direct recombination 
channel, in photodetector or solar cell applications.

\section{M\lowercase{o}S$_2$ with M\lowercase{o}S\lowercase{e}$_2$, M\lowercase{o}T\lowercase{e}$_2$, BN, and graphene}

\subsection{Electronic structure}

\begin{figure}[!ht]
\begin{center}
  \includegraphics[width=7.5cm]{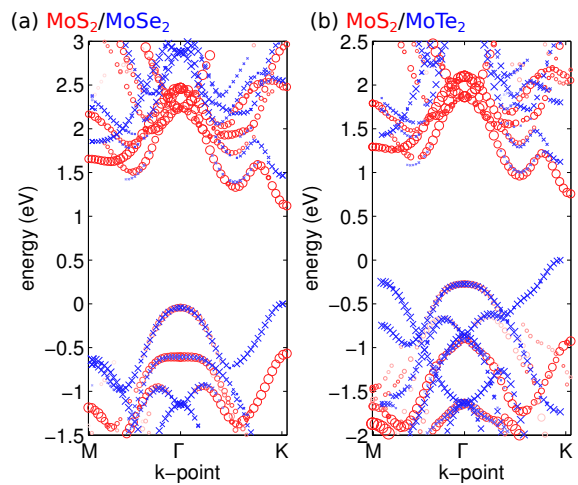}
\end{center}
\caption{\label{fig:MoSMoSebands}
(Color online)
Band structures of (a) MoS$_2$/MoSe$_2$ and (b) MoS$_2$/MoTe$_2$
heterostructures. 
Circles (red) denote projections to MoS$_2$ layer (opacity)
and to the corresponding reciprocal space directions (size of the marker).
Crosses (blue) denote similarly projections to MoSe$_2$ (a)
and MoTe$_2$ (b) layers.
}
\end{figure}

We next study electronic structure of the incommensurate systems.
The band structures of MoS$_2$/MoSe$_2$ and MoS$_2$/MoTe$_2$ are shown
in Fig.~\ref{fig:MoSMoSebands}(a,b). The states from the supercell calculation
are projected to the corresponding primitive cells of each layer.
Similar to the MoS$_2$/WS$_2$ system, the K-point states retain
their monolayer character, but the $\Gamma$-point states are split
with a small mixing of the wave function character.
Band structure of MoS$_2$/MoTe$_2$ looks somewhat different, 
as the MoS$_2$ VBM couples to the second highest VBM state of MoTe$_2$.
In both cases, the gap is direct in a sense that
both the VBM and CBM are located at the K-point,
although again these states are localized in different monolayers
in a type-II alignment.
We note, that the layer distance calculated with the PBE-D and 
PW86R-VV10 functionals again agree, both yielding 6.65 {\AA}.

\begin{figure}[!ht]
\begin{center}
  \includegraphics[width=7.5cm]{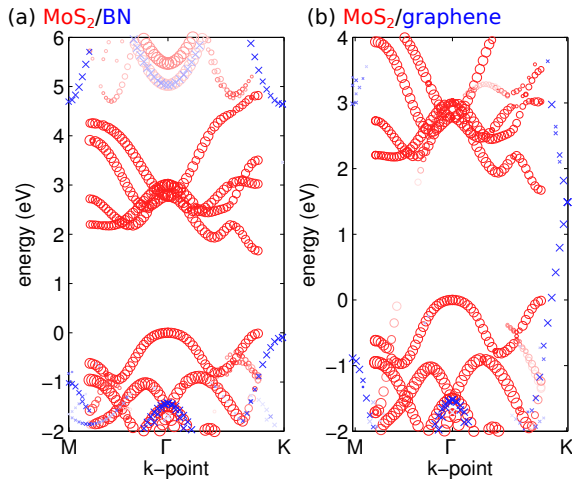}
\end{center}
\caption{\label{fig:MoSBNbands}
(Color online)
Band structures of (a) MoS$_2$/BN and (b) MoS$_2$/graphene
heterostructures. Notation as in Fig.~\ref{fig:MoSMoSebands}.
For each layer, the band structure is plotted up to its first 
Brillouin zone boundary. Due to different lattice constants,
this boundary is located at different wavevectors.
}
\end{figure}

The band structure of MoS$_2$/BN is shown in Fig.~\ref{fig:MoSBNbands}(a).
In this heterostructure, there are small corrugations in the BN layer:
the layer distance is largest for B atoms on top of Mo, and smallest
for B atoms on top of the center-of-hexagon of MoS$_2$.
However, the effect is small and does not affect the band structure.
The BN layer has practically no effect on the $\Gamma$-point
band edge of MoS$_2$. The $\Gamma$- and K-point edges are energetically
close and similar to that found for monolayer MoS$_2$.
In these calculations the K-point was below the
$\Gamma$-point by 15 meV, but inclusion of spin-orbit coupling
is expected to push VBM at K-point above the $\Gamma$-point valley.
The VBM of BN is close to, but below of, the VBM of MoS$_2$,
thereby yielding a type-I alignment.

The band structure of MoS$_2$/graphene is shown
in Fig.~\ref{fig:MoSBNbands}(b). Similar to BN, the interaction
with MoS$_2$ states around the $\Gamma$-point, as well as around K-point,
is negligible. The Dirac-point of graphene is slightly below CBM 
of MoS$_2$ in agreement with previous study.\cite{Ma11_Nanos}

\begin{figure*}[!ht]
\begin{center}
  \includegraphics[width=16cm]{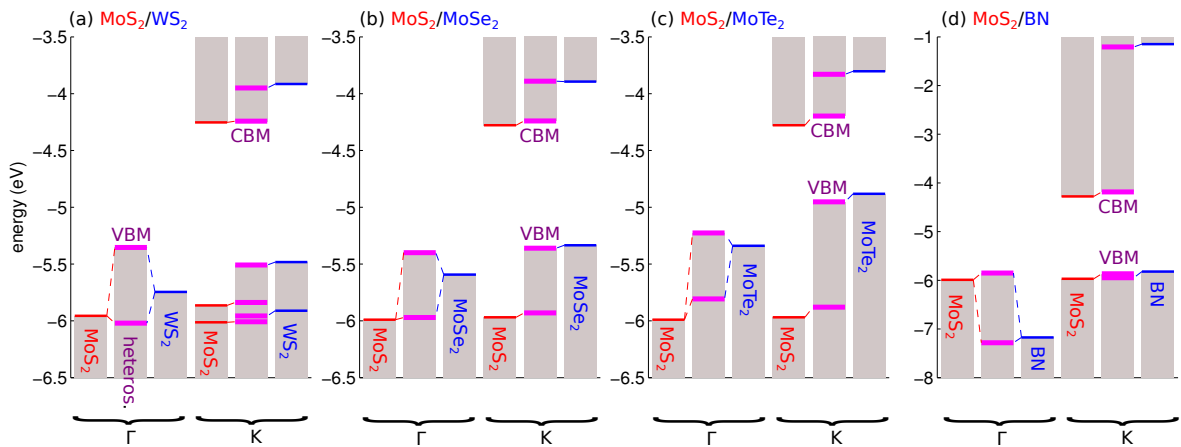}
\end{center}
\caption{\label{fig:levels}
(Color online)
Diagrams illustrating the interaction of band edge states around 
$\Gamma$- and K-points as the heterostructure is constructed from
the respective monolayers. The energies are given with respect
to the vacuum level. 
}
\end{figure*}

In order to understand the effects of interlayer interactions
more generally,
the band edge positions from all considered semiconducting
heterostructure systems are collected
in Fig.~\ref{fig:levels} together with the corresponding edges from
the respective monolayer systems.
Several trends can be observed:
(i) For TMD/TMD heterostructures, the nature of the VBM wave functions
around $\Gamma$-point leads to strongly split states. The magnitude
of the splitting depends on the position of the states prior to the
construction of the heterostructure. This behavior is akin to that
found in bilayer TMD systems or even in simple diatomic molecules.
We cannot directly probe the sensitivity of splitting on orientation,
but as it is observed in all systems here, we expect
the orientation to have small effect on the splitting.
(ii) Both the VBM and CBM states around K-point are consistently very 
close to those of the respective monolayers. As discussed above,
this is due to the wave functions being confined in the transition
metal sublattice. The small shifts of 0.1--0.2 eV are caused by
formation of interface dipole between the layers.
(iii) BN and graphene work well to ``insulate'' the TMD monolayer
so that their electronic structures remains very similar to the 
isolated layers, also around the $\Gamma$-point.

The TMD/TMD heterostructures examined so far all yielded type-II
alignment, which may be advantageous in separating electron/hole
pairs. On the other hand, in order to maximize photoemission, 
VBM and CBM should both reside at K-point
and localized to the same monolayer (i.e., type-I alignment).
Among the structures studied so far, MoS$_2$/BN was the only one
fulfilling these conditions. This should also hold for other
TMD/BN heterostructures.
In addition, using the above trends and the position of band edges
in isolated monolayers \cite{Jiang12_JPCC,Kang13_APL},
the following systems are also expected to show type-I alignment:
MoTe$_2$/WSe$_2$, MoSe$_2$/WS$_2$, HfS$_2$/ZrSe$_2$.

Naturally, the monolayer-monolayer interactions discussed within
this work are also applicable when substrate-monolayer
interactions are considered.
For instance, MoS$_2$ on BN substrate should still retain 
electronic structure very similar to that of a monolayer MoS$_2$. 
In fact, such system was found to work very well for optical
studies in Ref.~\onlinecite{Mak12_NNano}.
In case of TMD substrates, or more generally whenever
there is strong interaction at $\Gamma$-point with the substrate,
MoS$_2$ will easily become indirect gap due to the 
close energies of the valence band maxima at $\Gamma$- and K-points,
whereas WS$_2$, MoSe$_2$, or MoTe$_2$ are expected to retain 
their direct gaps.


\subsection{Optical properties}

The position of the band edges discussed above is naturally of 
importance for the optical properties as far as the valley 
population is concerned.
The incommensurate nature of the heterostructures considered here
also introduces a new issue of mismatch in valley location in
the reciprocal space. That is, the wave vector corresponding to
the K-point of the two monolayers have different length and orientation. 
For many of the TMD/TMD structures, the lengths are reasonably close, 
but the orientation of the layers is unknown.
If the heterostructure is constructed by manually placing monolayers 
on top of each other,\cite{Lee13_NL} the orientation is likely 
to be dominated by the deposition process and the layers 
do not reorient after that. If the heterostructure is constructed 
during growth or restacking of layers exfoliated in liquid,
\cite{Coleman11_Sci} the orientation might be determined by 
the total energy of stacking.

Since the K-points are aligned in MoS$_2$/WS$_2$, this system
is expected to show the strongest mixing in the optical transitions
and the mixing should become smaller in TMD/TMD systems where K-points 
are misaligned.
In the cases of TMD/BN and TMD/graphene, there is very little
mixing of the states around the band edges.
Since the transitions were seen to be largely decoupled 
in the MoS$_2$/WS$_2$ system, 
it therefore seems well justified to conclude that for all
heterostructures examined in this work the optical absorption 
spectrum should be well approximated by the sum of its 
monolayer constituents.

When interlayer excitons are considered, the misalignment of 
K-points would lead to indirect interlayer excitons.
Although physically interesting,
carrying out full $GW$+BSE calculations on these systems
is computationally demanding, and thus beyond the scope
of this paper.


\section{Conclusions}

A set of transition metal dichalcogenide heterostructures
were studied via first-principles calculations.
The adopted computational scheme enables us to study unstrained 
systems, thereby moving beyond the artificially strained or 
naturally commensurate systems considered previously.
With regards to the electronic structure, we find that 
the VBM at $\Gamma$-point is very sensitive to interlayer
interaction in TMD/TMD heterostructures. Consequently this 
determines whether VBM is located at the $\Gamma$ or K-point.
On the other hand, heterostructures with graphene or BN show 
only negligible interaction.
Furthermore, we calculated and analyzed the optical transitions
in the MoS$_2$/WS$_2$ system. The optical properties were found
to be only very weakly affected by the interlayer interactions. 
Novel optical characteristics of monolayer TMDs should be largely 
retained even when stacked with other layered materials, which 
should prove useful for amplifying their optical response. 
However, engineering the optical gap through heterostructuring
does not appear to work.
Optically inactive interlayer excitons were also found that
are of interest in light harvesting applications.
We hope our results
will guide and motivate future experiments in constructing layered 
structures and studying their unique optoelectronic properties.

\begin{acknowledgments}
We acknowledge financial support by the Academy of Finland through 
projects 218545 and 263416, the University of Helsinki Funds.
We also thank CSC Finland for generous grants of computer time.
\end{acknowledgments}


\begin{thebibliography}{53}%
\makeatletter
\providecommand \@ifxundefined [1]{%
 \@ifx{#1\undefined}
}%
\providecommand \@ifnum [1]{%
 \ifnum #1\expandafter \@firstoftwo
 \else \expandafter \@secondoftwo
 \fi
}%
\providecommand \@ifx [1]{%
 \ifx #1\expandafter \@firstoftwo
 \else \expandafter \@secondoftwo
 \fi
}%
\providecommand \natexlab [1]{#1}%
\providecommand \enquote  [1]{``#1''}%
\providecommand \bibnamefont  [1]{#1}%
\providecommand \bibfnamefont [1]{#1}%
\providecommand \citenamefont [1]{#1}%
\providecommand \href@noop [0]{\@secondoftwo}%
\providecommand \href [0]{\begingroup \@sanitize@url \@href}%
\providecommand \@href[1]{\@@startlink{#1}\@@href}%
\providecommand \@@href[1]{\endgroup#1\@@endlink}%
\providecommand \@sanitize@url [0]{\catcode `\\12\catcode `\$12\catcode
  `\&12\catcode `\#12\catcode `\^12\catcode `\_12\catcode `\%12\relax}%
\providecommand \@@startlink[1]{}%
\providecommand \@@endlink[0]{}%
\providecommand \url  [0]{\begingroup\@sanitize@url \@url }%
\providecommand \@url [1]{\endgroup\@href {#1}{\urlprefix }}%
\providecommand \urlprefix  [0]{URL }%
\providecommand \Eprint [0]{\href }%
\providecommand \doibase [0]{http://dx.doi.org/}%
\providecommand \selectlanguage [0]{\@gobble}%
\providecommand \bibinfo  [0]{\@secondoftwo}%
\providecommand \bibfield  [0]{\@secondoftwo}%
\providecommand \translation [1]{[#1]}%
\providecommand \BibitemOpen [0]{}%
\providecommand \bibitemStop [0]{}%
\providecommand \bibitemNoStop [0]{.\EOS\space}%
\providecommand \EOS [0]{\spacefactor3000\relax}%
\providecommand \BibitemShut  [1]{\csname bibitem#1\endcsname}%
\let\auto@bib@innerbib\@empty
\bibitem [{\citenamefont {Wilson}\ and\ \citenamefont
  {Yoffe}(1969)}]{Wilson69_AP}%
  \BibitemOpen
  \bibfield  {author} {\bibinfo {author} {\bibfnamefont {J.}~\bibnamefont
  {Wilson}}\ and\ \bibinfo {author} {\bibfnamefont {A.}~\bibnamefont {Yoffe}},\
  }\href {\doibase 10.1080/00018736900101307} {\bibfield  {journal} {\bibinfo
  {journal} {Adv. Phys.}\ }\textbf {\bibinfo {volume} {18}},\ \bibinfo {pages}
  {193} (\bibinfo {year} {1969})}\BibitemShut {NoStop}%
\bibitem [{\citenamefont {Tenne}\ \emph {et~al.}(1992)\citenamefont {Tenne},
  \citenamefont {Margulis}, \citenamefont {Genut},\ and\ \citenamefont
  {Hodes}}]{Tenne92_Nat}%
  \BibitemOpen
  \bibfield  {author} {\bibinfo {author} {\bibfnamefont {R.}~\bibnamefont
  {Tenne}}, \bibinfo {author} {\bibfnamefont {L.}~\bibnamefont {Margulis}},
  \bibinfo {author} {\bibfnamefont {M.}~\bibnamefont {Genut}}, \ and\ \bibinfo
  {author} {\bibfnamefont {G.}~\bibnamefont {Hodes}},\ }\href {\doibase
  10.1038/360444a0} {\bibfield  {journal} {\bibinfo  {journal} {Nature}\
  }\textbf {\bibinfo {volume} {360}},\ \bibinfo {pages} {444} (\bibinfo {year}
  {1992})}\BibitemShut {NoStop}%
\bibitem [{\citenamefont {Radisavljevic}\ \emph {et~al.}(2011)\citenamefont
  {Radisavljevic}, \citenamefont {Radenovic}, \citenamefont {Brivio},
  \citenamefont {Giacometti},\ and\ \citenamefont
  {Kis}}]{Radisavljevic11_NNano}%
  \BibitemOpen
  \bibfield  {author} {\bibinfo {author} {\bibfnamefont {B.}~\bibnamefont
  {Radisavljevic}}, \bibinfo {author} {\bibfnamefont {A.}~\bibnamefont
  {Radenovic}}, \bibinfo {author} {\bibfnamefont {J.}~\bibnamefont {Brivio}},
  \bibinfo {author} {\bibfnamefont {V.}~\bibnamefont {Giacometti}}, \ and\
  \bibinfo {author} {\bibfnamefont {A.}~\bibnamefont {Kis}},\ }\href {\doibase
  10.1038/nnano.2010.279} {\bibfield  {journal} {\bibinfo  {journal} {Nat.
  Nanot.}\ }\textbf {\bibinfo {volume} {6}},\ \bibinfo {pages} {147} (\bibinfo
  {year} {2011})}\BibitemShut {NoStop}%
\bibitem [{\citenamefont {Mak}\ \emph {et~al.}(2010)\citenamefont {Mak},
  \citenamefont {Lee}, \citenamefont {Hone}, \citenamefont {Shan},\ and\
  \citenamefont {Heinz}}]{Mak10_PRL}%
  \BibitemOpen
  \bibfield  {author} {\bibinfo {author} {\bibfnamefont {K.~F.}\ \bibnamefont
  {Mak}}, \bibinfo {author} {\bibfnamefont {C.}~\bibnamefont {Lee}}, \bibinfo
  {author} {\bibfnamefont {J.}~\bibnamefont {Hone}}, \bibinfo {author}
  {\bibfnamefont {J.}~\bibnamefont {Shan}}, \ and\ \bibinfo {author}
  {\bibfnamefont {T.~F.}\ \bibnamefont {Heinz}},\ }\href {\doibase
  10.1103/PhysRevLett.105.136805} {\bibfield  {journal} {\bibinfo  {journal}
  {Phys. Rev. Lett.}\ }\textbf {\bibinfo {volume} {105}},\ \bibinfo {pages}
  {136805} (\bibinfo {year} {2010})}\BibitemShut {NoStop}%
\bibitem [{\citenamefont {Xiao}\ \emph {et~al.}(2012)\citenamefont {Xiao},
  \citenamefont {Liu}, \citenamefont {Feng}, \citenamefont {Xu},\ and\
  \citenamefont {Yao}}]{Xiao12_PRL}%
  \BibitemOpen
  \bibfield  {author} {\bibinfo {author} {\bibfnamefont {D.}~\bibnamefont
  {Xiao}}, \bibinfo {author} {\bibfnamefont {G.-B.}\ \bibnamefont {Liu}},
  \bibinfo {author} {\bibfnamefont {W.}~\bibnamefont {Feng}}, \bibinfo {author}
  {\bibfnamefont {X.}~\bibnamefont {Xu}}, \ and\ \bibinfo {author}
  {\bibfnamefont {W.}~\bibnamefont {Yao}},\ }\href {\doibase
  10.1103/PhysRevLett.108.196802} {\bibfield  {journal} {\bibinfo  {journal}
  {Phys. Rev. Lett.}\ }\textbf {\bibinfo {volume} {108}},\ \bibinfo {pages}
  {196802} (\bibinfo {year} {2012})}\BibitemShut {NoStop}%
\bibitem [{\citenamefont {Mak}\ \emph {et~al.}(2013)\citenamefont {Mak},
  \citenamefont {He}, \citenamefont {Lee}, \citenamefont {Lee}, \citenamefont
  {Hone}, \citenamefont {Heinz},\ and\ \citenamefont {Shan}}]{Mak13_NMat}%
  \BibitemOpen
  \bibfield  {author} {\bibinfo {author} {\bibfnamefont {K.~F.}\ \bibnamefont
  {Mak}}, \bibinfo {author} {\bibfnamefont {K.}~\bibnamefont {He}}, \bibinfo
  {author} {\bibfnamefont {C.}~\bibnamefont {Lee}}, \bibinfo {author}
  {\bibfnamefont {G.~H.}\ \bibnamefont {Lee}}, \bibinfo {author} {\bibfnamefont
  {J.}~\bibnamefont {Hone}}, \bibinfo {author} {\bibfnamefont {T.~F.}\
  \bibnamefont {Heinz}}, \ and\ \bibinfo {author} {\bibfnamefont
  {J.}~\bibnamefont {Shan}},\ }\href {\doibase 10.1038/nmat3505} {\bibfield
  {journal} {\bibinfo  {journal} {Nat. Mater.}\ }\textbf {\bibinfo {volume}
  {12}},\ \bibinfo {pages} {207} (\bibinfo {year} {2013})}\BibitemShut
  {NoStop}%
\bibitem [{\citenamefont {Kaplan-Ashiri}\ \emph {et~al.}(2006)\citenamefont
  {Kaplan-Ashiri}, \citenamefont {Cohen}, \citenamefont {Gartsman},
  \citenamefont {Ivanovskaya}, \citenamefont {Heine}, \citenamefont {Seifert},
  \citenamefont {Wiesel}, \citenamefont {Wagner},\ and\ \citenamefont
  {Tenne}}]{KaplanAshiri06_PNAS}%
  \BibitemOpen
  \bibfield  {author} {\bibinfo {author} {\bibfnamefont {I.}~\bibnamefont
  {Kaplan-Ashiri}}, \bibinfo {author} {\bibfnamefont {S.~R.}\ \bibnamefont
  {Cohen}}, \bibinfo {author} {\bibfnamefont {K.}~\bibnamefont {Gartsman}},
  \bibinfo {author} {\bibfnamefont {V.}~\bibnamefont {Ivanovskaya}}, \bibinfo
  {author} {\bibfnamefont {T.}~\bibnamefont {Heine}}, \bibinfo {author}
  {\bibfnamefont {G.}~\bibnamefont {Seifert}}, \bibinfo {author} {\bibfnamefont
  {I.}~\bibnamefont {Wiesel}}, \bibinfo {author} {\bibfnamefont {H.~D.}\
  \bibnamefont {Wagner}}, \ and\ \bibinfo {author} {\bibfnamefont
  {R.}~\bibnamefont {Tenne}},\ }\href {\doibase 10.1073/pnas.0505640103}
  {\bibfield  {journal} {\bibinfo  {journal} {Proc. Natl. Acad. Sci. U.S.A.}\
  }\textbf {\bibinfo {volume} {103}},\ \bibinfo {pages} {523} (\bibinfo {year}
  {2006})}\BibitemShut {NoStop}%
\bibitem [{\citenamefont {Bertolazzi}\ \emph {et~al.}(2011)\citenamefont
  {Bertolazzi}, \citenamefont {Brivio},\ and\ \citenamefont
  {Kis}}]{Bertolazzi11_ACSNano}%
  \BibitemOpen
  \bibfield  {author} {\bibinfo {author} {\bibfnamefont {S.}~\bibnamefont
  {Bertolazzi}}, \bibinfo {author} {\bibfnamefont {J.}~\bibnamefont {Brivio}},
  \ and\ \bibinfo {author} {\bibfnamefont {A.}~\bibnamefont {Kis}},\ }\href
  {\doibase 10.1021/nn203879f} {\bibfield  {journal} {\bibinfo  {journal} {ACS
  Nano}\ }\textbf {\bibinfo {volume} {5}},\ \bibinfo {pages} {9703} (\bibinfo
  {year} {2011})}\BibitemShut {NoStop}%
\bibitem [{\citenamefont {Pu}\ \emph {et~al.}(2012)\citenamefont {Pu},
  \citenamefont {Yomogida}, \citenamefont {Liu}, \citenamefont {Li},
  \citenamefont {Iwasa},\ and\ \citenamefont {Takenobu}}]{Pu12_NL}%
  \BibitemOpen
  \bibfield  {author} {\bibinfo {author} {\bibfnamefont {J.}~\bibnamefont
  {Pu}}, \bibinfo {author} {\bibfnamefont {Y.}~\bibnamefont {Yomogida}},
  \bibinfo {author} {\bibfnamefont {K.-K.}\ \bibnamefont {Liu}}, \bibinfo
  {author} {\bibfnamefont {L.-J.}\ \bibnamefont {Li}}, \bibinfo {author}
  {\bibfnamefont {Y.}~\bibnamefont {Iwasa}}, \ and\ \bibinfo {author}
  {\bibfnamefont {T.}~\bibnamefont {Takenobu}},\ }\href {\doibase
  10.1021/nl301335q} {\bibfield  {journal} {\bibinfo  {journal} {Nano Lett.}\
  }\textbf {\bibinfo {volume} {12}},\ \bibinfo {pages} {4013} (\bibinfo {year}
  {2012})}\BibitemShut {NoStop}%
\bibitem [{\citenamefont {Novoselov}\ and\ \citenamefont
  {Castro~Neto}(2012)}]{Novoselov12_PS}%
  \BibitemOpen
  \bibfield  {author} {\bibinfo {author} {\bibfnamefont {K.~S.}\ \bibnamefont
  {Novoselov}}\ and\ \bibinfo {author} {\bibfnamefont {A.~H.}\ \bibnamefont
  {Castro~Neto}},\ }\href {\doibase 10.1088/0031-8949/2012/T146/014006}
  {\bibfield  {journal} {\bibinfo  {journal} {Physica Scripta}\ }\textbf
  {\bibinfo {volume} {T146}},\ \bibinfo {pages} {014006} (\bibinfo {year}
  {2012})}\BibitemShut {NoStop}%
\bibitem [{\citenamefont {Wang}\ \emph {et~al.}(2012)\citenamefont {Wang},
  \citenamefont {Kalantar-Zadeh}, \citenamefont {Kis}, \citenamefont
  {Coleman},\ and\ \citenamefont {Strano}}]{Wang12_NNano}%
  \BibitemOpen
  \bibfield  {author} {\bibinfo {author} {\bibfnamefont {Q.~H.}\ \bibnamefont
  {Wang}}, \bibinfo {author} {\bibfnamefont {K.}~\bibnamefont
  {Kalantar-Zadeh}}, \bibinfo {author} {\bibfnamefont {A.}~\bibnamefont {Kis}},
  \bibinfo {author} {\bibfnamefont {J.~N.}\ \bibnamefont {Coleman}}, \ and\
  \bibinfo {author} {\bibfnamefont {M.~S.}\ \bibnamefont {Strano}},\ }\href
  {\doibase 10.1038/nnano.2012.193} {\bibfield  {journal} {\bibinfo  {journal}
  {Nat. Nanot.}\ }\textbf {\bibinfo {volume} {7}},\ \bibinfo {pages} {699}
  (\bibinfo {year} {2012})}\BibitemShut {NoStop}%
\bibitem [{\citenamefont {Britnell}\ \emph {et~al.}(2012)\citenamefont
  {Britnell}, \citenamefont {Gorbachev}, \citenamefont {Jalil}, \citenamefont
  {Belle}, \citenamefont {Schedin}, \citenamefont {Mishchenko}, \citenamefont
  {Georgiou}, \citenamefont {Katsnelson}, \citenamefont {Eaves}, \citenamefont
  {Morozov}, \citenamefont {Peres}, \citenamefont {Leist}, \citenamefont
  {Geim}, \citenamefont {Novoselov},\ and\ \citenamefont
  {Ponomarenko}}]{Britnell12_Sci}%
  \BibitemOpen
  \bibfield  {author} {\bibinfo {author} {\bibfnamefont {L.}~\bibnamefont
  {Britnell}}, \bibinfo {author} {\bibfnamefont {R.~V.}\ \bibnamefont
  {Gorbachev}}, \bibinfo {author} {\bibfnamefont {R.}~\bibnamefont {Jalil}},
  \bibinfo {author} {\bibfnamefont {B.~D.}\ \bibnamefont {Belle}}, \bibinfo
  {author} {\bibfnamefont {F.}~\bibnamefont {Schedin}}, \bibinfo {author}
  {\bibfnamefont {A.}~\bibnamefont {Mishchenko}}, \bibinfo {author}
  {\bibfnamefont {T.}~\bibnamefont {Georgiou}}, \bibinfo {author}
  {\bibfnamefont {M.~I.}\ \bibnamefont {Katsnelson}}, \bibinfo {author}
  {\bibfnamefont {L.}~\bibnamefont {Eaves}}, \bibinfo {author} {\bibfnamefont
  {S.~V.}\ \bibnamefont {Morozov}}, \bibinfo {author} {\bibfnamefont
  {N.~M.~R.}\ \bibnamefont {Peres}}, \bibinfo {author} {\bibfnamefont
  {J.}~\bibnamefont {Leist}}, \bibinfo {author} {\bibfnamefont {A.~K.}\
  \bibnamefont {Geim}}, \bibinfo {author} {\bibfnamefont {K.~S.}\ \bibnamefont
  {Novoselov}}, \ and\ \bibinfo {author} {\bibfnamefont {L.~A.}\ \bibnamefont
  {Ponomarenko}},\ }\href {\doibase 10.1126/science.1218461} {\bibfield
  {journal} {\bibinfo  {journal} {Science}\ }\textbf {\bibinfo {volume}
  {335}},\ \bibinfo {pages} {947} (\bibinfo {year} {2012})}\BibitemShut
  {NoStop}%
\bibitem [{\citenamefont {Yu}\ \emph {et~al.}(2013)\citenamefont {Yu},
  \citenamefont {Li}, \citenamefont {Zhou}, \citenamefont {Chen}, \citenamefont
  {Wang}, \citenamefont {Huang},\ and\ \citenamefont {Duan}}]{Yu13_NMat}%
  \BibitemOpen
  \bibfield  {author} {\bibinfo {author} {\bibfnamefont {W.~J.}\ \bibnamefont
  {Yu}}, \bibinfo {author} {\bibfnamefont {Z.}~\bibnamefont {Li}}, \bibinfo
  {author} {\bibfnamefont {H.}~\bibnamefont {Zhou}}, \bibinfo {author}
  {\bibfnamefont {Y.}~\bibnamefont {Chen}}, \bibinfo {author} {\bibfnamefont
  {Y.}~\bibnamefont {Wang}}, \bibinfo {author} {\bibfnamefont {Y.}~\bibnamefont
  {Huang}}, \ and\ \bibinfo {author} {\bibfnamefont {X.}~\bibnamefont {Duan}},\
  }\href {\doibase 10.1038/nmat3518} {\bibfield  {journal} {\bibinfo  {journal}
  {Nat. Mater.}\ }\textbf {\bibinfo {volume} {12}},\ \bibinfo {pages} {246}
  (\bibinfo {year} {2013})}\BibitemShut {NoStop}%
\bibitem [{\citenamefont {Bertolazzi}\ \emph {et~al.}(2013)\citenamefont
  {Bertolazzi}, \citenamefont {Krasnozhon},\ and\ \citenamefont
  {Kis}}]{Bertolazzi13_ACSNano}%
  \BibitemOpen
  \bibfield  {author} {\bibinfo {author} {\bibfnamefont {S.}~\bibnamefont
  {Bertolazzi}}, \bibinfo {author} {\bibfnamefont {D.}~\bibnamefont
  {Krasnozhon}}, \ and\ \bibinfo {author} {\bibfnamefont {A.}~\bibnamefont
  {Kis}},\ }\href {\doibase 10.1021/nn3059136} {\bibfield  {journal} {\bibinfo
  {journal} {ACS Nano}\ }\textbf {\bibinfo {volume} {7}},\ \bibinfo {pages}
  {3246} (\bibinfo {year} {2013})}\BibitemShut {NoStop}%
\bibitem [{\citenamefont {Georgiou}\ \emph {et~al.}(2013)\citenamefont
  {Georgiou}, \citenamefont {Jalil}, \citenamefont {Belle}, \citenamefont
  {Britnell}, \citenamefont {Gorbachev}, \citenamefont {Morozov}, \citenamefont
  {Kim}, \citenamefont {Gholinia}, \citenamefont {Haigh}, \citenamefont
  {Makarovsky}, \citenamefont {Eaves}, \citenamefont {Ponomarenko},
  \citenamefont {Geim}, \citenamefont {Novoselov},\ and\ \citenamefont
  {Mishchenko}}]{Georgiou13_NNano}%
  \BibitemOpen
  \bibfield  {author} {\bibinfo {author} {\bibfnamefont {T.}~\bibnamefont
  {Georgiou}}, \bibinfo {author} {\bibfnamefont {R.}~\bibnamefont {Jalil}},
  \bibinfo {author} {\bibfnamefont {B.~D.}\ \bibnamefont {Belle}}, \bibinfo
  {author} {\bibfnamefont {L.}~\bibnamefont {Britnell}}, \bibinfo {author}
  {\bibfnamefont {R.~V.}\ \bibnamefont {Gorbachev}}, \bibinfo {author}
  {\bibfnamefont {S.~V.}\ \bibnamefont {Morozov}}, \bibinfo {author}
  {\bibfnamefont {Y.-J.}\ \bibnamefont {Kim}}, \bibinfo {author} {\bibfnamefont
  {A.}~\bibnamefont {Gholinia}}, \bibinfo {author} {\bibfnamefont {S.~J.}\
  \bibnamefont {Haigh}}, \bibinfo {author} {\bibfnamefont {O.}~\bibnamefont
  {Makarovsky}}, \bibinfo {author} {\bibfnamefont {L.}~\bibnamefont {Eaves}},
  \bibinfo {author} {\bibfnamefont {L.~A.}\ \bibnamefont {Ponomarenko}},
  \bibinfo {author} {\bibfnamefont {A.~K.}\ \bibnamefont {Geim}}, \bibinfo
  {author} {\bibfnamefont {K.~S.}\ \bibnamefont {Novoselov}}, \ and\ \bibinfo
  {author} {\bibfnamefont {A.}~\bibnamefont {Mishchenko}},\ }\href {\doibase
  10.1038/nnano.2012.224} {\bibfield  {journal} {\bibinfo  {journal} {Nat.
  Nanot.}\ }\textbf {\bibinfo {volume} {8}},\ \bibinfo {pages} {100} (\bibinfo
  {year} {2013})}\BibitemShut {NoStop}%
\bibitem [{\citenamefont {Komsa}\ \emph {et~al.}(2012)\citenamefont {Komsa},
  \citenamefont {Kotakoski}, \citenamefont {Kurasch}, \citenamefont {Lehtinen},
  \citenamefont {Kaiser},\ and\ \citenamefont {Krasheninnikov}}]{Komsa12_PRL}%
  \BibitemOpen
  \bibfield  {author} {\bibinfo {author} {\bibfnamefont {H.-P.}\ \bibnamefont
  {Komsa}}, \bibinfo {author} {\bibfnamefont {J.}~\bibnamefont {Kotakoski}},
  \bibinfo {author} {\bibfnamefont {S.}~\bibnamefont {Kurasch}}, \bibinfo
  {author} {\bibfnamefont {O.}~\bibnamefont {Lehtinen}}, \bibinfo {author}
  {\bibfnamefont {U.}~\bibnamefont {Kaiser}}, \ and\ \bibinfo {author}
  {\bibfnamefont {A.~V.}\ \bibnamefont {Krasheninnikov}},\ }\href {\doibase
  10.1103/PhysRevLett.109.035503} {\bibfield  {journal} {\bibinfo  {journal}
  {Phys. Rev. Lett.}\ }\textbf {\bibinfo {volume} {109}},\ \bibinfo {pages}
  {035503} (\bibinfo {year} {2012})}\BibitemShut {NoStop}%
\bibitem [{\citenamefont {Komsa}\ \emph {et~al.}(2013)\citenamefont {Komsa},
  \citenamefont {Kurasch}, \citenamefont {Lehtinen}, \citenamefont {Kaiser},\
  and\ \citenamefont {Krasheninnikov}}]{Komsa13_PRB}%
  \BibitemOpen
  \bibfield  {author} {\bibinfo {author} {\bibfnamefont {H.-P.}\ \bibnamefont
  {Komsa}}, \bibinfo {author} {\bibfnamefont {S.}~\bibnamefont {Kurasch}},
  \bibinfo {author} {\bibfnamefont {O.}~\bibnamefont {Lehtinen}}, \bibinfo
  {author} {\bibfnamefont {U.}~\bibnamefont {Kaiser}}, \ and\ \bibinfo {author}
  {\bibfnamefont {A.~V.}\ \bibnamefont {Krasheninnikov}},\ }\href {\doibase
  10.1103/PhysRevB.88.035301} {\bibfield  {journal} {\bibinfo  {journal} {Phys.
  Rev. B}\ }\textbf {\bibinfo {volume} {88}},\ \bibinfo {pages} {035301}
  (\bibinfo {year} {2013})}\BibitemShut {NoStop}%
\bibitem [{\citenamefont {Yadgarov}\ \emph {et~al.}(2012)\citenamefont
  {Yadgarov}, \citenamefont {Rosentsveig}, \citenamefont {Leitus},
  \citenamefont {Albu-Yaron}, \citenamefont {Moshkovich}, \citenamefont
  {Perfilyev}, \citenamefont {Vasic}, \citenamefont {Frenkel}, \citenamefont
  {Enyashin}, \citenamefont {Seifert}, \citenamefont {Rapoport},\ and\
  \citenamefont {Tenne}}]{Yadgarov12_ACIE}%
  \BibitemOpen
  \bibfield  {author} {\bibinfo {author} {\bibfnamefont {L.}~\bibnamefont
  {Yadgarov}}, \bibinfo {author} {\bibfnamefont {R.}~\bibnamefont
  {Rosentsveig}}, \bibinfo {author} {\bibfnamefont {G.}~\bibnamefont {Leitus}},
  \bibinfo {author} {\bibfnamefont {A.}~\bibnamefont {Albu-Yaron}}, \bibinfo
  {author} {\bibfnamefont {A.}~\bibnamefont {Moshkovich}}, \bibinfo {author}
  {\bibfnamefont {V.}~\bibnamefont {Perfilyev}}, \bibinfo {author}
  {\bibfnamefont {R.}~\bibnamefont {Vasic}}, \bibinfo {author} {\bibfnamefont
  {A.~I.}\ \bibnamefont {Frenkel}}, \bibinfo {author} {\bibfnamefont {A.~N.}\
  \bibnamefont {Enyashin}}, \bibinfo {author} {\bibfnamefont {G.}~\bibnamefont
  {Seifert}}, \bibinfo {author} {\bibfnamefont {L.}~\bibnamefont {Rapoport}}, \
  and\ \bibinfo {author} {\bibfnamefont {R.}~\bibnamefont {Tenne}},\ }\href
  {\doibase 10.1002/anie.201105324} {\bibfield  {journal} {\bibinfo  {journal}
  {Angew. Chem. Int. Ed.}\ }\textbf {\bibinfo {volume} {51}},\ \bibinfo {pages}
  {1148} (\bibinfo {year} {2012})}\BibitemShut {NoStop}%
\bibitem [{\citenamefont {Cheng}\ \emph {et~al.}(2013)\citenamefont {Cheng},
  \citenamefont {Zhu}, \citenamefont {Mi}, \citenamefont {Guo},\ and\
  \citenamefont {Schwingenschl\"ogl}}]{Cheng13_PRB}%
  \BibitemOpen
  \bibfield  {author} {\bibinfo {author} {\bibfnamefont {Y.~C.}\ \bibnamefont
  {Cheng}}, \bibinfo {author} {\bibfnamefont {Z.~Y.}\ \bibnamefont {Zhu}},
  \bibinfo {author} {\bibfnamefont {W.~B.}\ \bibnamefont {Mi}}, \bibinfo
  {author} {\bibfnamefont {Z.~B.}\ \bibnamefont {Guo}}, \ and\ \bibinfo
  {author} {\bibfnamefont {U.}~\bibnamefont {Schwingenschl\"ogl}},\ }\href
  {\doibase 10.1103/PhysRevB.87.100401} {\bibfield  {journal} {\bibinfo
  {journal} {Phys. Rev. B}\ }\textbf {\bibinfo {volume} {87}},\ \bibinfo
  {pages} {100401} (\bibinfo {year} {2013})}\BibitemShut {NoStop}%
\bibitem [{\citenamefont {Komsa}\ and\ \citenamefont
  {Krasheninnikov}(2012{\natexlab{a}})}]{Komsa12_JPCL}%
  \BibitemOpen
  \bibfield  {author} {\bibinfo {author} {\bibfnamefont {H.-P.}\ \bibnamefont
  {Komsa}}\ and\ \bibinfo {author} {\bibfnamefont {A.~V.}\ \bibnamefont
  {Krasheninnikov}},\ }\href {\doibase 10.1021/jz301673x} {\bibfield  {journal}
  {\bibinfo  {journal} {J. Phys. Chem. Lett.}\ }\textbf {\bibinfo {volume}
  {3}},\ \bibinfo {pages} {3652} (\bibinfo {year}
  {2012}{\natexlab{a}})}\BibitemShut {NoStop}%
\bibitem [{\citenamefont {Kang}\ \emph
  {et~al.}(2013{\natexlab{a}})\citenamefont {Kang}, \citenamefont {Tongay},
  \citenamefont {Li},\ and\ \citenamefont {Wu}}]{Kang13_JAP}%
  \BibitemOpen
  \bibfield  {author} {\bibinfo {author} {\bibfnamefont {J.}~\bibnamefont
  {Kang}}, \bibinfo {author} {\bibfnamefont {S.}~\bibnamefont {Tongay}},
  \bibinfo {author} {\bibfnamefont {J.}~\bibnamefont {Li}}, \ and\ \bibinfo
  {author} {\bibfnamefont {J.}~\bibnamefont {Wu}},\ }\href {\doibase
  10.1063/1.4799126} {\bibfield  {journal} {\bibinfo  {journal} {J. Appl.
  Phys.}\ }\textbf {\bibinfo {volume} {113}},\ \bibinfo {eid} {143703}
  (\bibinfo {year} {2013}{\natexlab{a}})}\BibitemShut {NoStop}%
\bibitem [{\citenamefont {Dumcenco}\ \emph {et~al.}(2013)\citenamefont
  {Dumcenco}, \citenamefont {Kobayashi}, \citenamefont {Liu}, \citenamefont
  {Huang},\ and\ \citenamefont {Suenaga}}]{Dumcenco13_NComm}%
  \BibitemOpen
  \bibfield  {author} {\bibinfo {author} {\bibfnamefont {D.~O.}\ \bibnamefont
  {Dumcenco}}, \bibinfo {author} {\bibfnamefont {H.}~\bibnamefont {Kobayashi}},
  \bibinfo {author} {\bibfnamefont {Z.}~\bibnamefont {Liu}}, \bibinfo {author}
  {\bibfnamefont {Y.-S.}\ \bibnamefont {Huang}}, \ and\ \bibinfo {author}
  {\bibfnamefont {K.}~\bibnamefont {Suenaga}},\ }\href {\doibase
  10.1038/ncomms2351} {\bibfield  {journal} {\bibinfo  {journal} {Nat. Comm.}\
  }\textbf {\bibinfo {volume} {4}},\ \bibinfo {pages} {1351} (\bibinfo {year}
  {2013})}\BibitemShut {NoStop}%
\bibitem [{\citenamefont {Kuc}\ \emph {et~al.}(2011)\citenamefont {Kuc},
  \citenamefont {Zibouche},\ and\ \citenamefont {Heine}}]{Kuc11_PRB}%
  \BibitemOpen
  \bibfield  {author} {\bibinfo {author} {\bibfnamefont {A.}~\bibnamefont
  {Kuc}}, \bibinfo {author} {\bibfnamefont {N.}~\bibnamefont {Zibouche}}, \
  and\ \bibinfo {author} {\bibfnamefont {T.}~\bibnamefont {Heine}},\ }\href
  {\doibase 10.1103/PhysRevB.83.245213} {\bibfield  {journal} {\bibinfo
  {journal} {Phys. Rev. B}\ }\textbf {\bibinfo {volume} {83}},\ \bibinfo
  {pages} {245213} (\bibinfo {year} {2011})}\BibitemShut {NoStop}%
\bibitem [{\citenamefont {Ko\ifmmode~\acute{s}\else \'{s}\fi{}mider}\ and\
  \citenamefont {Fern\'andez-Rossier}(2013)}]{Kosmider13_PRB}%
  \BibitemOpen
  \bibfield  {author} {\bibinfo {author} {\bibfnamefont {K.}~\bibnamefont
  {Ko\ifmmode~\acute{s}\else \'{s}\fi{}mider}}\ and\ \bibinfo {author}
  {\bibfnamefont {J.}~\bibnamefont {Fern\'andez-Rossier}},\ }\href {\doibase
  10.1103/PhysRevB.87.075451} {\bibfield  {journal} {\bibinfo  {journal} {Phys.
  Rev. B}\ }\textbf {\bibinfo {volume} {87}},\ \bibinfo {pages} {075451}
  (\bibinfo {year} {2013})}\BibitemShut {NoStop}%
\bibitem [{\citenamefont {Terrones}\ \emph {et~al.}(2013)\citenamefont
  {Terrones}, \citenamefont {L\'opez-Ur\'ias},\ and\ \citenamefont
  {Terrones}}]{Terrones13_SRep}%
  \BibitemOpen
  \bibfield  {author} {\bibinfo {author} {\bibfnamefont {H.}~\bibnamefont
  {Terrones}}, \bibinfo {author} {\bibfnamefont {F.}~\bibnamefont
  {L\'opez-Ur\'ias}}, \ and\ \bibinfo {author} {\bibfnamefont {M.}~\bibnamefont
  {Terrones}},\ }\href {\doibase 10.1038/srep01549} {\bibfield  {journal}
  {\bibinfo  {journal} {Scientific Reports}\ }\textbf {\bibinfo {volume} {3}},\
  \bibinfo {pages} {1549} (\bibinfo {year} {2013})}\BibitemShut {NoStop}%
\bibitem [{\citenamefont {Kou}\ \emph {et~al.}(2013)\citenamefont {Kou},
  \citenamefont {Frauenheim},\ and\ \citenamefont {Chen}}]{Kou13_JPCL}%
  \BibitemOpen
  \bibfield  {author} {\bibinfo {author} {\bibfnamefont {L.}~\bibnamefont
  {Kou}}, \bibinfo {author} {\bibfnamefont {T.}~\bibnamefont {Frauenheim}}, \
  and\ \bibinfo {author} {\bibfnamefont {C.}~\bibnamefont {Chen}},\ }\href
  {\doibase 10.1021/jz400668d} {\bibfield  {journal} {\bibinfo  {journal} {The
  Journal of Physical Chemistry Letters}\ }\textbf {\bibinfo {volume} {4}},\
  \bibinfo {pages} {1730} (\bibinfo {year} {2013})}\BibitemShut {NoStop}%
\bibitem [{\citenamefont {Shi}\ \emph {et~al.}(2012)\citenamefont {Shi},
  \citenamefont {Zhou}, \citenamefont {Lu}, \citenamefont {Fang}, \citenamefont
  {Lee}, \citenamefont {Hsu}, \citenamefont {Kim}, \citenamefont {Kim},
  \citenamefont {Yang}, \citenamefont {Li}, \citenamefont {Idrobo},\ and\
  \citenamefont {Kong}}]{Shi12_NL}%
  \BibitemOpen
  \bibfield  {author} {\bibinfo {author} {\bibfnamefont {Y.}~\bibnamefont
  {Shi}}, \bibinfo {author} {\bibfnamefont {W.}~\bibnamefont {Zhou}}, \bibinfo
  {author} {\bibfnamefont {A.-Y.}\ \bibnamefont {Lu}}, \bibinfo {author}
  {\bibfnamefont {W.}~\bibnamefont {Fang}}, \bibinfo {author} {\bibfnamefont
  {Y.-H.}\ \bibnamefont {Lee}}, \bibinfo {author} {\bibfnamefont {A.~L.}\
  \bibnamefont {Hsu}}, \bibinfo {author} {\bibfnamefont {S.~M.}\ \bibnamefont
  {Kim}}, \bibinfo {author} {\bibfnamefont {K.~K.}\ \bibnamefont {Kim}},
  \bibinfo {author} {\bibfnamefont {H.~Y.}\ \bibnamefont {Yang}}, \bibinfo
  {author} {\bibfnamefont {L.-J.}\ \bibnamefont {Li}}, \bibinfo {author}
  {\bibfnamefont {J.-C.}\ \bibnamefont {Idrobo}}, \ and\ \bibinfo {author}
  {\bibfnamefont {J.}~\bibnamefont {Kong}},\ }\href {\doibase
  10.1021/nl204562j} {\bibfield  {journal} {\bibinfo  {journal} {Nano Lett.}\
  }\textbf {\bibinfo {volume} {12}},\ \bibinfo {pages} {2784} (\bibinfo {year}
  {2012})}\BibitemShut {NoStop}%
\bibitem [{\citenamefont {Yun}\ \emph {et~al.}(2012)\citenamefont {Yun},
  \citenamefont {Han}, \citenamefont {Hong}, \citenamefont {Kim},\ and\
  \citenamefont {Lee}}]{Yun12_PRB}%
  \BibitemOpen
  \bibfield  {author} {\bibinfo {author} {\bibfnamefont {W.~S.}\ \bibnamefont
  {Yun}}, \bibinfo {author} {\bibfnamefont {S.~W.}\ \bibnamefont {Han}},
  \bibinfo {author} {\bibfnamefont {S.~C.}\ \bibnamefont {Hong}}, \bibinfo
  {author} {\bibfnamefont {I.~G.}\ \bibnamefont {Kim}}, \ and\ \bibinfo
  {author} {\bibfnamefont {J.~D.}\ \bibnamefont {Lee}},\ }\href {\doibase
  10.1103/PhysRevB.85.033305} {\bibfield  {journal} {\bibinfo  {journal} {Phys.
  Rev. B}\ }\textbf {\bibinfo {volume} {85}},\ \bibinfo {pages} {033305}
  (\bibinfo {year} {2012})}\BibitemShut {NoStop}%
\bibitem [{\citenamefont {Johari}\ and\ \citenamefont
  {Shenoy}(2012)}]{Johari12_ACSNano}%
  \BibitemOpen
  \bibfield  {author} {\bibinfo {author} {\bibfnamefont {P.}~\bibnamefont
  {Johari}}\ and\ \bibinfo {author} {\bibfnamefont {V.~B.}\ \bibnamefont
  {Shenoy}},\ }\href {\doibase 10.1021/nn301320r} {\bibfield  {journal}
  {\bibinfo  {journal} {ACS Nano}\ }\textbf {\bibinfo {volume} {6}},\ \bibinfo
  {pages} {5449} (\bibinfo {year} {2012})}\BibitemShut {NoStop}%
\bibitem [{\citenamefont {Shi}\ \emph {et~al.}(2013)\citenamefont {Shi},
  \citenamefont {Pan}, \citenamefont {Zhang},\ and\ \citenamefont
  {Yakobson}}]{Shi13_PRB}%
  \BibitemOpen
  \bibfield  {author} {\bibinfo {author} {\bibfnamefont {H.}~\bibnamefont
  {Shi}}, \bibinfo {author} {\bibfnamefont {H.}~\bibnamefont {Pan}}, \bibinfo
  {author} {\bibfnamefont {Y.-W.}\ \bibnamefont {Zhang}}, \ and\ \bibinfo
  {author} {\bibfnamefont {B.~I.}\ \bibnamefont {Yakobson}},\ }\href {\doibase
  10.1103/PhysRevB.87.155304} {\bibfield  {journal} {\bibinfo  {journal} {Phys.
  Rev. B}\ }\textbf {\bibinfo {volume} {87}},\ \bibinfo {pages} {155304}
  (\bibinfo {year} {2013})}\BibitemShut {NoStop}%
\bibitem [{\citenamefont {Kresse}\ and\ \citenamefont {Hafner}(1993)}]{kres1}%
  \BibitemOpen
  \bibfield  {author} {\bibinfo {author} {\bibfnamefont {G.}~\bibnamefont
  {Kresse}}\ and\ \bibinfo {author} {\bibfnamefont {J.}~\bibnamefont
  {Hafner}},\ }\href@noop {} {\bibfield  {journal} {\bibinfo  {journal} {Phys.
  Rev. B}\ }\textbf {\bibinfo {volume} {47}},\ \bibinfo {pages} {558} (\bibinfo
  {year} {1993})}\BibitemShut {NoStop}%
\bibitem [{\citenamefont {Kresse}\ and\ \citenamefont
  {Furthm{\"{u}}ller}(1996)}]{kres2}%
  \BibitemOpen
  \bibfield  {author} {\bibinfo {author} {\bibfnamefont {G.}~\bibnamefont
  {Kresse}}\ and\ \bibinfo {author} {\bibfnamefont {J.}~\bibnamefont
  {Furthm{\"{u}}ller}},\ }\href@noop {} {\bibfield  {journal} {\bibinfo
  {journal} {Comput. Mat. Sci.}\ }\textbf {\bibinfo {volume} {6}},\ \bibinfo
  {pages} {15} (\bibinfo {year} {1996})}\BibitemShut {NoStop}%
\bibitem [{\citenamefont {Perdew}\ \emph {et~al.}(1996)\citenamefont {Perdew},
  \citenamefont {Burke},\ and\ \citenamefont {Ernzerhof}}]{Perdew96}%
  \BibitemOpen
  \bibfield  {author} {\bibinfo {author} {\bibfnamefont {J.~P.}\ \bibnamefont
  {Perdew}}, \bibinfo {author} {\bibfnamefont {K.}~\bibnamefont {Burke}}, \
  and\ \bibinfo {author} {\bibfnamefont {M.}~\bibnamefont {Ernzerhof}},\
  }\href@noop {} {\bibfield  {journal} {\bibinfo  {journal} {Phys. Rev. Lett.}\
  }\textbf {\bibinfo {volume} {77}},\ \bibinfo {pages} {3865} (\bibinfo {year}
  {1996})}\BibitemShut {NoStop}%
\bibitem [{\citenamefont {Grimme}(2006)}]{Grimme06}%
  \BibitemOpen
  \bibfield  {author} {\bibinfo {author} {\bibfnamefont {S.}~\bibnamefont
  {Grimme}},\ }\href {\doibase 10.1002/jcc.20495} {\bibfield  {journal}
  {\bibinfo  {journal} {J. Comput. Chem.}\ }\textbf {\bibinfo {volume} {27}},\
  \bibinfo {pages} {1787} (\bibinfo {year} {2006})}\BibitemShut {NoStop}%
\bibitem [{\citenamefont {Vydrov}\ and\ \citenamefont
  {Van~Voorhis}(2010)}]{Vydrov10_JCP}%
  \BibitemOpen
  \bibfield  {author} {\bibinfo {author} {\bibfnamefont {O.~A.}\ \bibnamefont
  {Vydrov}}\ and\ \bibinfo {author} {\bibfnamefont {T.}~\bibnamefont
  {Van~Voorhis}},\ }\href@noop {} {\bibfield  {journal} {\bibinfo  {journal}
  {J. Chem. Phys.}\ }\textbf {\bibinfo {volume} {133}},\ \bibinfo {pages}
  {244103} (\bibinfo {year} {2010})}\BibitemShut {NoStop}%
\bibitem [{\citenamefont {Bj\"orkman}(2012)}]{Bjorkman12_PRB}%
  \BibitemOpen
  \bibfield  {author} {\bibinfo {author} {\bibfnamefont {T.}~\bibnamefont
  {Bj\"orkman}},\ }\href@noop {} {\bibfield  {journal} {\bibinfo  {journal}
  {Phys. Rev. B}\ }\textbf {\bibinfo {volume} {86}},\ \bibinfo {pages} {165109}
  (\bibinfo {year} {2012})}\BibitemShut {NoStop}%
\bibitem [{\citenamefont {Bj\"orkman}\ \emph {et~al.}(2012)\citenamefont
  {Bj\"orkman}, \citenamefont {Gulans}, \citenamefont {Krasheninnikov},\ and\
  \citenamefont {Nieminen}}]{Bjorkman12_PRL}%
  \BibitemOpen
  \bibfield  {author} {\bibinfo {author} {\bibfnamefont {T.}~\bibnamefont
  {Bj\"orkman}}, \bibinfo {author} {\bibfnamefont {A.}~\bibnamefont {Gulans}},
  \bibinfo {author} {\bibfnamefont {A.~V.}\ \bibnamefont {Krasheninnikov}}, \
  and\ \bibinfo {author} {\bibfnamefont {R.~M.}\ \bibnamefont {Nieminen}},\
  }\href {\doibase 10.1103/PhysRevLett.108.235502} {\bibfield  {journal}
  {\bibinfo  {journal} {Phys. Rev. Lett.}\ }\textbf {\bibinfo {volume} {108}},\
  \bibinfo {pages} {235502} (\bibinfo {year} {2012})}\BibitemShut {NoStop}%
\bibitem [{\citenamefont {Popescu}\ and\ \citenamefont
  {Zunger}(2010)}]{Popescu10_PRL}%
  \BibitemOpen
  \bibfield  {author} {\bibinfo {author} {\bibfnamefont {V.}~\bibnamefont
  {Popescu}}\ and\ \bibinfo {author} {\bibfnamefont {A.}~\bibnamefont
  {Zunger}},\ }\href@noop {} {\bibfield  {journal} {\bibinfo  {journal} {Phys.
  Rev. Lett.}\ }\textbf {\bibinfo {volume} {104}},\ \bibinfo {pages} {236403}
  (\bibinfo {year} {2010})}\BibitemShut {NoStop}%
\bibitem [{\citenamefont {Komsa}\ and\ \citenamefont
  {Krasheninnikov}(2012{\natexlab{b}})}]{Komsa12_PRB2}%
  \BibitemOpen
  \bibfield  {author} {\bibinfo {author} {\bibfnamefont {H.-P.}\ \bibnamefont
  {Komsa}}\ and\ \bibinfo {author} {\bibfnamefont {A.~V.}\ \bibnamefont
  {Krasheninnikov}},\ }\href {\doibase 10.1103/PhysRevB.86.241201} {\bibfield
  {journal} {\bibinfo  {journal} {Phys. Rev. B}\ }\textbf {\bibinfo {volume}
  {86}},\ \bibinfo {pages} {241201} (\bibinfo {year}
  {2012}{\natexlab{b}})}\BibitemShut {NoStop}%
\bibitem [{\citenamefont {Shishkin}\ and\ \citenamefont
  {Kresse}(2006)}]{Shishkin06_PRB}%
  \BibitemOpen
  \bibfield  {author} {\bibinfo {author} {\bibfnamefont {M.}~\bibnamefont
  {Shishkin}}\ and\ \bibinfo {author} {\bibfnamefont {G.}~\bibnamefont
  {Kresse}},\ }\href {\doibase 10.1103/PhysRevB.74.035101} {\bibfield
  {journal} {\bibinfo  {journal} {Phys. Rev. B}\ }\textbf {\bibinfo {volume}
  {74}},\ \bibinfo {pages} {035101} (\bibinfo {year} {2006})}\BibitemShut
  {NoStop}%
\bibitem [{\citenamefont {Rohlfing}\ and\ \citenamefont
  {Louie}(1998)}]{Rohlfing98_PRL}%
  \BibitemOpen
  \bibfield  {author} {\bibinfo {author} {\bibfnamefont {M.}~\bibnamefont
  {Rohlfing}}\ and\ \bibinfo {author} {\bibfnamefont {S.~G.}\ \bibnamefont
  {Louie}},\ }\href@noop {} {\bibfield  {journal} {\bibinfo  {journal} {Phys.
  Rev. Lett.}\ }\textbf {\bibinfo {volume} {81}},\ \bibinfo {pages} {2312}
  (\bibinfo {year} {1998})}\BibitemShut {NoStop}%
\bibitem [{\citenamefont {Rinke}\ \emph {et~al.}(2005)\citenamefont {Rinke},
  \citenamefont {Qteish}, \citenamefont {Neugebauer}, \citenamefont
  {Freysoldt},\ and\ \citenamefont {Scheffler}}]{Rinke05_NJP}%
  \BibitemOpen
  \bibfield  {author} {\bibinfo {author} {\bibfnamefont {P.}~\bibnamefont
  {Rinke}}, \bibinfo {author} {\bibfnamefont {A.}~\bibnamefont {Qteish}},
  \bibinfo {author} {\bibfnamefont {J.}~\bibnamefont {Neugebauer}}, \bibinfo
  {author} {\bibfnamefont {C.}~\bibnamefont {Freysoldt}}, \ and\ \bibinfo
  {author} {\bibfnamefont {M.}~\bibnamefont {Scheffler}},\ }\href {\doibase
  10.1088/1367-2630/7/1/126} {\bibfield  {journal} {\bibinfo  {journal} {New J.
  Phys.}\ }\textbf {\bibinfo {volume} {7}},\ \bibinfo {pages} {126} (\bibinfo
  {year} {2005})}\BibitemShut {NoStop}%
\bibitem [{\citenamefont {Bechstedt}\ \emph {et~al.}(2009)\citenamefont
  {Bechstedt}, \citenamefont {Fuchs},\ and\ \citenamefont
  {Kresse}}]{Bechstedt09_PSSb}%
  \BibitemOpen
  \bibfield  {author} {\bibinfo {author} {\bibfnamefont {F.}~\bibnamefont
  {Bechstedt}}, \bibinfo {author} {\bibfnamefont {F.}~\bibnamefont {Fuchs}}, \
  and\ \bibinfo {author} {\bibfnamefont {G.}~\bibnamefont {Kresse}},\ }\href
  {\doibase 10.1002/pssb.200945074} {\bibfield  {journal} {\bibinfo  {journal}
  {phys. stat. sol. (b)}\ }\textbf {\bibinfo {volume} {246}},\ \bibinfo {pages}
  {1877} (\bibinfo {year} {2009})}\BibitemShut {NoStop}%
\bibitem [{\citenamefont {Wirtz}\ \emph {et~al.}(2006)\citenamefont {Wirtz},
  \citenamefont {Marini},\ and\ \citenamefont {Rubio}}]{Wirtz06_PRL}%
  \BibitemOpen
  \bibfield  {author} {\bibinfo {author} {\bibfnamefont {L.}~\bibnamefont
  {Wirtz}}, \bibinfo {author} {\bibfnamefont {A.}~\bibnamefont {Marini}}, \
  and\ \bibinfo {author} {\bibfnamefont {A.}~\bibnamefont {Rubio}},\ }\href
  {\doibase 10.1103/PhysRevLett.96.126104} {\bibfield  {journal} {\bibinfo
  {journal} {Phys. Rev. Lett.}\ }\textbf {\bibinfo {volume} {96}},\ \bibinfo
  {pages} {126104} (\bibinfo {year} {2006})}\BibitemShut {NoStop}%
\bibitem [{\citenamefont {Molina-S\'anchez}\ \emph {et~al.}(2013)\citenamefont
  {Molina-S\'anchez}, \citenamefont {Sangalli}, \citenamefont {Hummer},
  \citenamefont {Marini},\ and\ \citenamefont {Wirtz}}]{MolinaSanchez13_PRB}%
  \BibitemOpen
  \bibfield  {author} {\bibinfo {author} {\bibfnamefont {A.}~\bibnamefont
  {Molina-S\'anchez}}, \bibinfo {author} {\bibfnamefont {D.}~\bibnamefont
  {Sangalli}}, \bibinfo {author} {\bibfnamefont {K.}~\bibnamefont {Hummer}},
  \bibinfo {author} {\bibfnamefont {A.}~\bibnamefont {Marini}}, \ and\ \bibinfo
  {author} {\bibfnamefont {L.}~\bibnamefont {Wirtz}},\ }\href {\doibase
  10.1103/PhysRevB.88.045412} {\bibfield  {journal} {\bibinfo  {journal} {Phys.
  Rev. B}\ }\textbf {\bibinfo {volume} {88}},\ \bibinfo {pages} {045412}
  (\bibinfo {year} {2013})}\BibitemShut {NoStop}%
\bibitem [{\citenamefont {Jiang}(2012)}]{Jiang12_JPCC}%
  \BibitemOpen
  \bibfield  {author} {\bibinfo {author} {\bibfnamefont {H.}~\bibnamefont
  {Jiang}},\ }\href {\doibase 10.1021/jp300079d} {\bibfield  {journal}
  {\bibinfo  {journal} {J. Phys. Chem. C}\ }\textbf {\bibinfo {volume} {116}},\
  \bibinfo {pages} {7664} (\bibinfo {year} {2012})}\BibitemShut {NoStop}%
\bibitem [{\citenamefont {Kang}\ \emph
  {et~al.}(2013{\natexlab{b}})\citenamefont {Kang}, \citenamefont {Tongay},
  \citenamefont {Zhou}, \citenamefont {Li},\ and\ \citenamefont
  {Wu}}]{Kang13_APL}%
  \BibitemOpen
  \bibfield  {author} {\bibinfo {author} {\bibfnamefont {J.}~\bibnamefont
  {Kang}}, \bibinfo {author} {\bibfnamefont {S.}~\bibnamefont {Tongay}},
  \bibinfo {author} {\bibfnamefont {J.}~\bibnamefont {Zhou}}, \bibinfo {author}
  {\bibfnamefont {J.}~\bibnamefont {Li}}, \ and\ \bibinfo {author}
  {\bibfnamefont {J.}~\bibnamefont {Wu}},\ }\href {\doibase 10.1063/1.4774090}
  {\bibfield  {journal} {\bibinfo  {journal} {Appl. Phys. Lett.}\ }\textbf
  {\bibinfo {volume} {102}},\ \bibinfo {eid} {012111} (\bibinfo {year}
  {2013}{\natexlab{b}})}\BibitemShut {NoStop}%
\bibitem [{\citenamefont {Zhao}\ \emph {et~al.}(2013)\citenamefont {Zhao},
  \citenamefont {Ghorannevis}, \citenamefont {Chu}, \citenamefont {Toh},
  \citenamefont {Kloc}, \citenamefont {Tan},\ and\ \citenamefont
  {Eda}}]{Zhao13_ACSNano}%
  \BibitemOpen
  \bibfield  {author} {\bibinfo {author} {\bibfnamefont {W.}~\bibnamefont
  {Zhao}}, \bibinfo {author} {\bibfnamefont {Z.}~\bibnamefont {Ghorannevis}},
  \bibinfo {author} {\bibfnamefont {L.}~\bibnamefont {Chu}}, \bibinfo {author}
  {\bibfnamefont {M.}~\bibnamefont {Toh}}, \bibinfo {author} {\bibfnamefont
  {C.}~\bibnamefont {Kloc}}, \bibinfo {author} {\bibfnamefont {P.-H.}\
  \bibnamefont {Tan}}, \ and\ \bibinfo {author} {\bibfnamefont
  {G.}~\bibnamefont {Eda}},\ }\href {\doibase 10.1021/nn305275h} {\bibfield
  {journal} {\bibinfo  {journal} {ACS Nano}\ }\textbf {\bibinfo {volume} {7}},\
  \bibinfo {pages} {791} (\bibinfo {year} {2013})}\BibitemShut {NoStop}%
\bibitem [{\citenamefont {Mak}\ \emph {et~al.}(2012)\citenamefont {Mak},
  \citenamefont {He}, \citenamefont {Shan},\ and\ \citenamefont
  {Heinz}}]{Mak12_NNano}%
  \BibitemOpen
  \bibfield  {author} {\bibinfo {author} {\bibfnamefont {K.~F.}\ \bibnamefont
  {Mak}}, \bibinfo {author} {\bibfnamefont {K.}~\bibnamefont {He}}, \bibinfo
  {author} {\bibfnamefont {J.}~\bibnamefont {Shan}}, \ and\ \bibinfo {author}
  {\bibfnamefont {T.~F.}\ \bibnamefont {Heinz}},\ }\href {\doibase
  10.1038/nnano.2012.96} {\bibfield  {journal} {\bibinfo  {journal} {Nat.
  Nanot.}\ }\textbf {\bibinfo {volume} {7}},\ \bibinfo {pages} {494} (\bibinfo
  {year} {2012})}\BibitemShut {NoStop}%
\bibitem [{\citenamefont {Guti\'errez}\ \emph {et~al.}(2013)\citenamefont
  {Guti\'errez}, \citenamefont {Perea-L\'opez}, \citenamefont {El\'ias},
  \citenamefont {Berkdemir}, \citenamefont {Wang}, \citenamefont {Lv},
  \citenamefont {L\'opez-Ur\'ias}, \citenamefont {Crespi}, \citenamefont
  {Terrones},\ and\ \citenamefont {Terrones}}]{Gutierrez13_NL}%
  \BibitemOpen
  \bibfield  {author} {\bibinfo {author} {\bibfnamefont {H.~R.}\ \bibnamefont
  {Guti\'errez}}, \bibinfo {author} {\bibfnamefont {N.}~\bibnamefont
  {Perea-L\'opez}}, \bibinfo {author} {\bibfnamefont {A.~L.}\ \bibnamefont
  {El\'ias}}, \bibinfo {author} {\bibfnamefont {A.}~\bibnamefont {Berkdemir}},
  \bibinfo {author} {\bibfnamefont {B.}~\bibnamefont {Wang}}, \bibinfo {author}
  {\bibfnamefont {R.}~\bibnamefont {Lv}}, \bibinfo {author} {\bibfnamefont
  {F.}~\bibnamefont {L\'opez-Ur\'ias}}, \bibinfo {author} {\bibfnamefont
  {V.~H.}\ \bibnamefont {Crespi}}, \bibinfo {author} {\bibfnamefont
  {H.}~\bibnamefont {Terrones}}, \ and\ \bibinfo {author} {\bibfnamefont
  {M.}~\bibnamefont {Terrones}},\ }\href {\doibase 10.1021/nl3026357}
  {\bibfield  {journal} {\bibinfo  {journal} {Nano Letters}\ }\textbf {\bibinfo
  {volume} {0}},\ \bibinfo {pages} {0} (\bibinfo {year} {2013})}\BibitemShut
  {NoStop}%
\bibitem [{\citenamefont {Ma}\ \emph {et~al.}(2011)\citenamefont {Ma},
  \citenamefont {Dai}, \citenamefont {Guo}, \citenamefont {Niu},\ and\
  \citenamefont {Huang}}]{Ma11_Nanos}%
  \BibitemOpen
  \bibfield  {author} {\bibinfo {author} {\bibfnamefont {Y.}~\bibnamefont
  {Ma}}, \bibinfo {author} {\bibfnamefont {Y.}~\bibnamefont {Dai}}, \bibinfo
  {author} {\bibfnamefont {M.}~\bibnamefont {Guo}}, \bibinfo {author}
  {\bibfnamefont {C.}~\bibnamefont {Niu}}, \ and\ \bibinfo {author}
  {\bibfnamefont {B.}~\bibnamefont {Huang}},\ }\href {\doibase
  10.1039/C1NR10577A} {\bibfield  {journal} {\bibinfo  {journal} {Nanoscale}\
  }\textbf {\bibinfo {volume} {3}},\ \bibinfo {pages} {3883} (\bibinfo {year}
  {2011})}\BibitemShut {NoStop}%
\bibitem [{\citenamefont {Lee}\ \emph {et~al.}(2013)\citenamefont {Lee},
  \citenamefont {Yu}, \citenamefont {Wang}, \citenamefont {Fang}, \citenamefont
  {Ling}, \citenamefont {Shi}, \citenamefont {Lin}, \citenamefont {Huang},
  \citenamefont {Chang}, \citenamefont {Chang}, \citenamefont {Dresselhaus},
  \citenamefont {Palacios}, \citenamefont {Li},\ and\ \citenamefont
  {Kong}}]{Lee13_NL}%
  \BibitemOpen
  \bibfield  {author} {\bibinfo {author} {\bibfnamefont {Y.-H.}\ \bibnamefont
  {Lee}}, \bibinfo {author} {\bibfnamefont {L.}~\bibnamefont {Yu}}, \bibinfo
  {author} {\bibfnamefont {H.}~\bibnamefont {Wang}}, \bibinfo {author}
  {\bibfnamefont {W.}~\bibnamefont {Fang}}, \bibinfo {author} {\bibfnamefont
  {X.}~\bibnamefont {Ling}}, \bibinfo {author} {\bibfnamefont {Y.}~\bibnamefont
  {Shi}}, \bibinfo {author} {\bibfnamefont {C.-T.}\ \bibnamefont {Lin}},
  \bibinfo {author} {\bibfnamefont {J.-K.}\ \bibnamefont {Huang}}, \bibinfo
  {author} {\bibfnamefont {M.-T.}\ \bibnamefont {Chang}}, \bibinfo {author}
  {\bibfnamefont {C.-S.}\ \bibnamefont {Chang}}, \bibinfo {author}
  {\bibfnamefont {M.}~\bibnamefont {Dresselhaus}}, \bibinfo {author}
  {\bibfnamefont {T.}~\bibnamefont {Palacios}}, \bibinfo {author}
  {\bibfnamefont {L.-J.}\ \bibnamefont {Li}}, \ and\ \bibinfo {author}
  {\bibfnamefont {J.}~\bibnamefont {Kong}},\ }\href {\doibase
  10.1021/nl400687n} {\bibfield  {journal} {\bibinfo  {journal} {Nano Lett.}\
  }\textbf {\bibinfo {volume} {13}},\ \bibinfo {pages} {1852} (\bibinfo {year}
  {2013})}\BibitemShut {NoStop}%
\bibitem [{\citenamefont {Coleman}\ \emph {et~al.}(2011)\citenamefont
  {Coleman}, \citenamefont {Lotya}, \citenamefont {O'Neill}, \citenamefont
  {Bergin}, \citenamefont {King}, \citenamefont {Khan}, \citenamefont {Young},
  \citenamefont {Gaucher}, \citenamefont {De}, \citenamefont {Smith},
  \citenamefont {Shvets}, \citenamefont {Arora}, \citenamefont {Stanton},
  \citenamefont {Kim}, \citenamefont {Lee}, \citenamefont {Kim}, \citenamefont
  {Duesberg}, \citenamefont {Hallam}, \citenamefont {Boland}, \citenamefont
  {Wang}, \citenamefont {Donegan}, \citenamefont {Grunlan}, \citenamefont
  {Moriarty}, \citenamefont {Shmeliov}, \citenamefont {Nicholls}, \citenamefont
  {Perkins}, \citenamefont {Grieveson}, \citenamefont {Theuwissen},
  \citenamefont {McComb}, \citenamefont {Nellist},\ and\ \citenamefont
  {Nicolosi}}]{Coleman11_Sci}%
  \BibitemOpen
  \bibfield  {author} {\bibinfo {author} {\bibfnamefont {J.~N.}\ \bibnamefont
  {Coleman}}, \bibinfo {author} {\bibfnamefont {M.}~\bibnamefont {Lotya}},
  \bibinfo {author} {\bibfnamefont {A.}~\bibnamefont {O'Neill}}, \bibinfo
  {author} {\bibfnamefont {S.~D.}\ \bibnamefont {Bergin}}, \bibinfo {author}
  {\bibfnamefont {P.~J.}\ \bibnamefont {King}}, \bibinfo {author}
  {\bibfnamefont {U.}~\bibnamefont {Khan}}, \bibinfo {author} {\bibfnamefont
  {K.}~\bibnamefont {Young}}, \bibinfo {author} {\bibfnamefont
  {A.}~\bibnamefont {Gaucher}}, \bibinfo {author} {\bibfnamefont
  {S.}~\bibnamefont {De}}, \bibinfo {author} {\bibfnamefont {R.~J.}\
  \bibnamefont {Smith}}, \bibinfo {author} {\bibfnamefont {I.~V.}\ \bibnamefont
  {Shvets}}, \bibinfo {author} {\bibfnamefont {S.~K.}\ \bibnamefont {Arora}},
  \bibinfo {author} {\bibfnamefont {G.}~\bibnamefont {Stanton}}, \bibinfo
  {author} {\bibfnamefont {H.-Y.}\ \bibnamefont {Kim}}, \bibinfo {author}
  {\bibfnamefont {K.}~\bibnamefont {Lee}}, \bibinfo {author} {\bibfnamefont
  {G.~T.}\ \bibnamefont {Kim}}, \bibinfo {author} {\bibfnamefont {G.~S.}\
  \bibnamefont {Duesberg}}, \bibinfo {author} {\bibfnamefont {T.}~\bibnamefont
  {Hallam}}, \bibinfo {author} {\bibfnamefont {J.~J.}\ \bibnamefont {Boland}},
  \bibinfo {author} {\bibfnamefont {J.~J.}\ \bibnamefont {Wang}}, \bibinfo
  {author} {\bibfnamefont {J.~F.}\ \bibnamefont {Donegan}}, \bibinfo {author}
  {\bibfnamefont {J.~C.}\ \bibnamefont {Grunlan}}, \bibinfo {author}
  {\bibfnamefont {G.}~\bibnamefont {Moriarty}}, \bibinfo {author}
  {\bibfnamefont {A.}~\bibnamefont {Shmeliov}}, \bibinfo {author}
  {\bibfnamefont {R.~J.}\ \bibnamefont {Nicholls}}, \bibinfo {author}
  {\bibfnamefont {J.~M.}\ \bibnamefont {Perkins}}, \bibinfo {author}
  {\bibfnamefont {E.~M.}\ \bibnamefont {Grieveson}}, \bibinfo {author}
  {\bibfnamefont {K.}~\bibnamefont {Theuwissen}}, \bibinfo {author}
  {\bibfnamefont {D.~W.}\ \bibnamefont {McComb}}, \bibinfo {author}
  {\bibfnamefont {P.~D.}\ \bibnamefont {Nellist}}, \ and\ \bibinfo {author}
  {\bibfnamefont {V.}~\bibnamefont {Nicolosi}},\ }\href {\doibase
  10.1126/science.1194975} {\bibfield  {journal} {\bibinfo  {journal}
  {Science}\ }\textbf {\bibinfo {volume} {331}},\ \bibinfo {pages} {568}
  (\bibinfo {year} {2011})}\BibitemShut {NoStop}%
\end{thebibliography}

%

\end{document}